\documentclass[12pt]{article}

\begin{document}

\begin{center}

{\Large {\bf String theory provides landmarks fixing \\ 

\vspace{0,5cm}

RS branes' positions: Possibility to deduce \\ 

\vspace{0,5cm}

large mass hierarchy from small numbers}}

\vspace{2cm}

{Boris L. Altshuler}\footnote[1]{E-mail adresses: altshuler@mtu-net.ru \& altshul@lpi.ru}

\vspace{0,5cm}

{\it Theoretical Physics Department, P.N. Lebedev Physical
Institute, \\  53 Leninsky Prospect, Moscow, 119991, Russia}

\vspace{2cm}

{\bf Abstract}

\end{center}

Two-branes RS-I 5-dimensional model is generalized in higher dimensional string-induced theory with dilaton and $n$-form field. It is supposed that "hidden" and "visible" Randall-Sundrum branes are located at the boundaries of region of space-time where low-energy supergravity description is valid. This permits to determine mass scale hierarchy which calculated value proves to be strongly dependent on dimensionalities of subspaces.

\newpage

\section{Introduction}
\qquad This paper develops the ideas of previous work~\cite{I&IIAltsh} where it was shown that large mass hierarchy may be calculated from reasonably small values of dimensionalities and dilaton coupling constant. The basic idea there was the introduction into brane's action of the unconventional mass term of $n$-form tensor field. However consistency of the approach demanded the negative sign of this term; that puts questions about stability of the model. Here we discard this idea and will try to show that large mass hierarchy may be calculated from naturally small numbers in frames of string-based supergravity theory which Bose sector in string metric in $D$-dimensional space-time is given by the action~\cite{Metsaev, Duff, Aharony}:

\begin{equation}
\label{1}
S^{(D)}=\frac{1}{2k^{2}}\int\Big\{e^{-2\Phi}\left(R_{\it{st}}+4(\nabla\Phi)^2+\alpha'R_{\it{st}}^{2}+\cdots\right)-\frac{1}{2n!}F_{(n)}^2\Big\}\sqrt{-g^{(D)}}\,d^{D}x.
\end{equation}
Below the notation $1/2k^{2}=M^{D-2}$ will be used, $M^{-1}$ and $\alpha'^{1/2}$ in (\ref{1}) are of order of string length; $R_{\it{st}}$ is curvature in $D$ dimensions
in string metric, $F_{(n)}$ is $n$-form tensor field, dots symbolize higher order terms in $\alpha'R_{\it{st}}$. The consideration of low energy effective action is justified only in case these terms are small; string coupling constant $e^{\Phi}$ must be small as well.
These conditions will be used below to single out the "permitted" region of space-time which in turn surves a tool to estimate the mass scale hierarchy number.

The ideology of calculation of mass scale hierarchy in this paper basically follows that of the Randall-Sundrum two-branes model (RS-I)~\cite{Randall} where space-time
is bounded by two co-dimension one branes with $Z_{2}$ symmetry imposed at each of them. Those are: "hidden" high energy brane and "visible" electro-weak scale brane where matter
of our Universe is trapped (the idea of trapping visible matter on a sub-manifold of space-time comes back to pioneer papers~\cite{Rubakov,Akama}). Mass hierarchy will be determined below as a product of two factors: ratio of values of warp factor at the positions of the RS branes and ratio of calculated value of Planck mass $M_{\rm{Pl}}$ to fundamental string scale $M$ (we'll see that in the model under consideration $M$ is some
intermediate scale between $M_{\rm{Pl}}$ and electroweek scale $m$). 

In string theory where field theory effective action reproduces the massless sector of the string S-matrix the very notion of space-time metric makes sense only in low energy
approximation. Thus the words "permitted region of space-time" should not be understood literally. The main hypotheses of this paper looks physically rather simple: we suppose that because of unknown quantum high order effects visible
matter of our universe is trapped at the boundary of existence of space-time where curvature reaches string scale. Calculations below will be
carried out for two options of this boundary condition written down for curvature in string metric $R_{\it{st}}$ (like in (\ref{1})) and for curvature in Einstein metric $R_{\it{E}}$ (like
in (\ref{2}) below) - see inequalities (\ref{11}), (\ref{12}) correspondingly.

\section{Description of the model}

\qquad The standard Einstein-Hilbert bulk action resulting from rescaling of action (\ref{1}) will be used in the paper:

\begin{eqnarray}
\label{2}
&&S^{(D)}_{\it{bulk}}=M^{D-2}\int\Big\{R_{\it{E}}-\frac{1}{2}(\nabla\phi)^2-\frac{1}{2n!}e^{\alpha\phi}F_{(n)}^2+  \nonumber  \\
&&+M^{-2}e^{-\eta\phi}R_{\it{E}}^{2}+\cdots \Big\}\sqrt{-g^{(D)}}\,d^{D}x+\rm{GH},
\end{eqnarray}
where "Planck mass" $M$ in $D$ dimensions is given by fundamental string scale, $R_{\it{E}}$ - scalar curvature in Einstein metric $g_{AB}$ in $D$ dimensions,
$\rm{GH}$ - Gibbons-Hawking surface term,
$\phi$ - dilaton field in Einstein metric coupled to the $n$-form field with a coupling constant $\alpha$. The rescaling of action (\ref{1}) gives following
values of dimensionless constants $\alpha$ and $\eta$ in (\ref{2}):

\begin{equation}
\label{3}
\alpha=\frac{D-2n}{\sqrt{2(D-2)}}, \qquad \eta=\frac{2}{\sqrt{2(D-2)}};
\end{equation}
these values of $\alpha$ incorporate in particular its values (1/2 and 1) for well known $F_{(4)}$ and $F_{(3)}$ terms of $D$10 type IIA supergravity~\cite{Duff, Aharony}. Higher order curvature terms in (\ref{2}) will not be taken into account in dynamical equations considered below; however they must be small and this condition is crucial for our calculations as it was pointed out in the Introduction.

The well known anzats for metric of $D$-dimensional space-time is taken:

\begin{equation}
\label{4}
ds_{(D)}^{2}=b^{2}ds^{2}_{(p+1)}+ a^{2}d\Omega_{(n)}+N^{2}dz^{2},
\end{equation}
where $D=p+n+2$, $ds^{2}_{(p+1)}$ is metric of $(p+1)$-dimensional flat space-time $M^{(p+1)}$, $d\Omega_{(n)}$ is metric of unit $n$-dimensional sphere. The antisymmetric $n$~-~form field is supposed to be given by magnetic monopole solution on sphere $S^{n}$ ($Q$ is $n$-form magnetic charge):

\begin{equation}
\label{5}
F_{(n)}=F_{i_{1}\cdots i_{n}}=Q\epsilon_{i_{1}\cdots i_{n}},
\end{equation}
wherefrom with account of (\ref{4}):

\begin{equation}
\label{6}
\frac{F_{(n)}^{2}}{n!}=\frac{Q^{2}}{a^{2n}}.
\end{equation}
\\

For the higher-dimensional space-time determined by metric (\ref{4}) the co-dimension one branes limiting space-time may be considered as $p$-branes of the RS-I model (where $p=3$) smeared (or "delocalized") over the surface of sphere $S^{n}$. Then fulfillment of Israel jump conditions demands anisotropy of brane's energy-momentum tensor. Since sphere has a non-zero curvature the most natural way to introduce this anisotropy is perhaps to introduce in the brane action the induced curvature term (see e.g. recent discussion in ~\cite{Mavromatos}). Thus we take the following brane action:

\begin{equation}
\label{7}
S_{\it{br}}=M^{D-2}\int\Big\{-\sigma e^{\gamma\phi}+\kappa R^{(D-1)}_{\it br} e^{\delta\phi}\Big\}\sqrt{-g^{(D-1)}}\,d^{D-1}x,
\end{equation}
where $\sigma$ and $\kappa$ are dimensional constants characterizing brain's tension and strength of brane's gravity (consistency of the model demands their fine-tuning); $\sqrt{-g^{(D-1)}}=b^{p+1}a^{n}$ (see (\ref{4})); for metric (\ref{4}) brane's curvature in (\ref{7}) is equal to curvature of $n$-sphere:

\begin{equation}
\label{8}
R^{(D-1)}_{\it br}=\frac{n(n-1)}{a^{2}};
\end{equation}

To determine dimensionless brane's dilaton coupling constants $\gamma$ and $\delta$ in (\ref{7}) we shall follow the idea of demanding the same scaling multiplicative behavior of bulk (\ref{2}) and brane (\ref{7}) actions under the scale transformation of $g_{AB}$ and $e^{\phi}$~\cite{Duff}. This gives unambiguously:

\begin{equation}
\label{9}
\frac{\gamma}{\alpha}=-\frac{1}{2(n-1)} \, , \qquad {}  \frac{\delta}{\alpha}=\frac{1}{2(n-1)}.
\end{equation}

With these "scaling" values of brane's dilaton coupling constants Israel boundary conditions in the theory given by the sum of the actions (\ref{2}), (\ref{7}) do not determine brane's position (like in RS model and contrary to the approach of ~\cite{I&IIAltsh}), but impose certain consistency relations between constants of the action and of the solution. We shall consider (\ref{7}) as the action of high-energy "hidden" brane; $\sigma$ and $\kappa$ in (\ref{7}) are positive for this brane.

The key issue is to determine positions of the "hidden" and "visible" branes. RS model has no tool for it, hence it is not predictive in calculation of mass scale hierarchy. Contrary to the RS-I where curvature is constant in the model at hand there are variable dilaton field and variable scalar curvature $R$. And we shall $\it{postulate}$ that slice of $D$-dimensional space-time "squeezed" between two co-dimension one branes is determined by two conditions of applicability of the low energy string approximation when supergravity description by effective actions (\ref{1}) or equivalently (\ref{2}) is valid. Those conditions are (see e.g. ~\cite{Aharony}, p.19):

1) Effective string coupling $e^{\phi}$ needs to be kept small:

\begin{equation}
\label{10}
e^{\phi}\le \,1,
\end{equation}
this condition determines boundary value $\phi=0$ at the "hidden" brane position (taken below at $z=0$).

2) Curvature must be small compared to the string scale:

\begin{equation}
\label{11}
|R_{\it{st}}|=|R_{\it{E}}| \, e^{-\eta\phi}\le \, M^{2},
\end{equation}
wherefrom the position of "visible" brane $z=z^{(\it{st})}_{\it{vis}}$ will be determined. For illustrative purposes another value of the "visible" brane position $z=z^{(\it{E})}_{\it{vis}}$ will be calculated from the analogous condition written for curvature $R_{\it{E}}$ in Einstein metric:

\begin{equation}
\label{12}
|R_{\it{E}}|\le\,M^{2}.
\end{equation}
This condition proves to be essentially less restrictive than (\ref{11}).

The knowledge of $z=z^{(\it{st})}_{\it{vis}}$ or $z=z^{(\it{E})}_{\it{vis}}$ will permit to calculate the first factor ($m^{(\it{st})}/M$ or $m^{(\it{E})}/M$) determining mass scale hierarchy for two cases described by inequalities (\ref{11}) or (\ref{12}).

\section{Dynamical equations and their solution}

\qquad Varying of the sum of bulk and brane actions (\ref{2}), (\ref{7}) with account of (\ref{6}), (\ref{8}) gives constraint equation and three second order equations for scales $N$, $b$, $a$ of metric (\ref{4}) and for dilaton field $\phi$ (prime means derivation over $z$):

\begin{eqnarray}
\label{13}
&&-\frac{n(n-1)}{2a^{2}}+\frac{1}{2N^{2}}\Bigg[p(p+1)\frac{b'^2}{b^2}+n(n-1)\frac{a'^2}{a^2}+2n(p+1)\frac{b'a'}{ba}\Bigg]= \nonumber \\
&&=-\frac{Q^2}{4a^{2n}}e^{\alpha\phi}+\frac{\phi'^2}{4N^2};
\end{eqnarray}
\\
\begin{eqnarray}
\label{14}
&&-\frac{1}{N^2}\Bigg[\frac{b''}{b}+\frac{b'}{b}\Bigg(-\frac{N'}{N}+p\frac{b'}{b}+n\frac{a'}{a}\Bigg)\Bigg]= \nonumber \\
&&=\frac{1}{2(D-2)}\,\Bigg[-\frac{Q^2}{a^{2n}} e^{\alpha\phi}\,(n-1)+\hat\sigma e^{\gamma\phi}+\hat\kappa\,\frac{n(n-1)}{a^2}e^{\delta\phi}\Bigg];
\end{eqnarray}
\\
\begin{eqnarray}
\label{15}
&&\frac{n-1}{a^2}-\frac{1}{N^2}\Bigg[\frac{a''}{a}+\frac{a'}{a}\Bigg(-\frac{N'}{N}+(p+1)\frac{b'}{b}+(n-1)\frac{a'}{a}\Bigg)\Bigg]= \nonumber
\\
&&=\frac{1}{2(D-2)}\,\Bigg[\frac{Q^2}{a^{2n}} e^{\alpha\phi}\,(p+1)+\hat\sigma e^{\gamma\phi}-\hat\kappa\,\frac{n(n-1)}{a^2}\,\frac{2p+n}{n}\,e^{\delta\phi}\Bigg];
\end{eqnarray}
\\
\begin{equation}
\label{16}
\frac{1}{J}\Bigg[\frac{J\phi'}{N^{2}}\Bigg]'=\alpha \,\frac{Q^2}{2a^{2n}} e^{\alpha\phi}+\gamma \,\hat\sigma e^{\gamma\phi}-\delta \,\hat\kappa\,\frac{n(n-1)}{a^2}e^{\delta\phi},
\end{equation}
\\
where

\begin{equation}
\label{17}
J\equiv b^{p+1}a^{n}N, \qquad \hat\sigma\equiv \sigma \frac{\delta(z-z_{\it{br}})}{N}, \qquad \hat\kappa\equiv \kappa \frac{\delta(z-z_{\it {br}})}{N}.
\end{equation}

The following bulk solution of Eq-s (\ref{13})-(\ref{16}) will be used below:

\begin{equation}
\label{18}
N=1, \quad b=\left(1-\frac{z}{L}\right)^{\xi}, \quad a=a_{0}\left(1-\frac{z}{L}\right), \quad e^{\alpha\phi}=\left(1-\frac{z}{L}\right)^{2(n-1)}, 
\end{equation}
\\
where

\begin{eqnarray}
\label{19}
&&\xi=\frac{2(n-1)^{2}}{(p+n)\alpha^{2}}, \qquad L=a_{0}\,\frac{[2(n-1)(p+1)+(p+n)\alpha^{2}]}{(p+n)\alpha^{2}}, \nonumber \\
&&\quad \frac{Q^{2}}{a_{0}^{2(n-1)}}=\frac{4(n-1)^{2}(p+n)}{2(n-1)(p+1)+(p+n)\alpha^{2}}.
\end{eqnarray}
\\
It is seen that singular point $z=L$ of the solution (\ref{18}) is given by value of $n$-sphere radius $a=a_{0}$ at $z=0$ (which is connected with $n$-form charge $Q$); $a_{0}$ in turn will be determined by Israel jump conditions through parameters of brane action (\ref{7}). Bulk solution (\ref{18}) is a near horizon asymptotic of the well known extremal $p$-brane solution. 

Shift in $z$ will just change multiplicatively $a$, $b$, $e^{\alpha\phi}$, $L$ in (\ref{18}) and thanks to "scaling" choice (\ref{9}) of brane's dilaton coupling constants will result after all in redefinition of $M$. This will change ratio $\sigma/M$ which in any case is an arbitrary constant of the theory given by sum of actions (\ref{2}), (\ref{7}). Thus the choice $\phi(0)=0$ in (\ref{18}) is always possible and is not an extra condition at all. The same point $z=0$, according to (\ref{10}), will be chosen the position of "hidden" high energy brane.

In the limit $\alpha\to 0$ (\ref{19}) gives $L \to \infty$ and solution (\ref{18}) as expected comes to standard Randall-Sundrum solution received by compactification of $D$-dimensional space-time to $AdS_{p+2}\times S^{n}$:

\begin{equation}
\label{20}
N=1, \quad b=e^{-Hz}, \quad H\equiv \frac{(n-1)}{(p+1)a_{0}}, \quad a=a_{0}, \quad \phi=1. 
\end{equation}
\\

We shall not put down Israel jump conditions which immediately follow from integration of Eq-s (\ref{14})-(\ref{16}) over brane's position with use of definitions (\ref{17}), bulk solution (\ref{18}) and $Z_{2}$-symmetry imposed at the brane: $\int b''=2b'(z=0)$. To sum up: because of "scaling" choice (\ref{9}) of brane's dilaton coupling constants three jump conditions are consistent, they do not include the position of the brane, but they connect parameters of bulk solution (\ref{18}) and brane action (\ref{7}):

\begin{equation}
\label{21}
a_{0} \sigma=\frac{2[2(n-1)^{2}(2p+n)+n(p+n)\alpha^{2}]}{2(n-1)(p+1)+(p+n)\alpha^{2}}\equiv K_{1}, 
\end{equation}
and they also demand fine-tuning of brane action constants $\sigma$ and $\kappa$:

\begin{equation}
\label{22}
\sigma \kappa=\frac{4 \, [2(n-1)^{2}(2p+n)+n(p+n)\alpha^{2}]{}\,\,[2(n-1)^{2}-(p+n)\alpha^{2}]}{(n-1)\,[2(n-1)(p+1)+(p+n)\alpha^{2}]^{2}}.
\end{equation}

The success of all the approach depends on fulfillment of the inequality (\ref{11}) (or (\ref{12})) in some slice $0<z<z_{\it{vis}}$ of space-time. For the model in consideration and with account of Einstein equations scalar curvature $R_{\it{E}}$ in (\ref{2}) is given by the expression:

\begin{equation}
\label{23}
R_{\it{E}}=\frac{D-2n}{D-2}\,\frac{Q^{2}}{2a^{2n}}\,e^{\alpha\phi}+\frac{\phi'^{2}}{2}.
\end{equation}
For the bulk solution (\ref{18}) this gives:

\begin{equation}
\label{24}
R_{\it{E}}=\frac{K_{2}}{a_{0}^{2}}\left(1-\frac{z}{L}\right)^{-2},
\end{equation}
where

\begin{equation}
\label{25}
K_{2}\equiv \frac{4(n-1)^{2}(p+1)\,[(p-n+2)(n-1)+(p+n)\alpha^{2}]}{[2(n-1)(p+1)+(p+n)\alpha^{2}]^{2}}.
\end{equation}

With use of (\ref{21}), (\ref{24}) inequalities (\ref{10}), (\ref{11}) give the permitted interval of proper coordinate $z$:

\begin{equation}
\label{26}
0<z<z^{(\it{st})}_{\it{vis}}= L\left(1-\left[\frac{|K_{2}|^{1/2}}{K_{1}}\frac{\sigma}{M}\right]^{1/\beta}\right), \qquad \beta\,\equiv\,1+(n-1)\frac{\eta}{\alpha},
\end{equation}
$\alpha$, $\eta$, $K_{1}$, $K_{2}$ are given in (\ref{3}), (\ref{21}) and (\ref{25}).
Alternatively (\ref{10}), (\ref{12}) will give:

\begin{equation}
\label{27}
0<z<z^{(\it{E})}_{\it{vis}}=L\left(1-\frac{|K_{2}|^{1/2}}{K_{1}}\frac{\sigma}{M}\right).
\end{equation}

\section{Calculation of Planck mass and mass scale hierarchy in 4 dimensions}

\qquad In what follows the general case of $p>3$ will be considered, hence the flat manifold $M^{(p+1)}$ in (\ref{4}) is supposed to be a product of 4-dimensional Lorentzian space-time $M^{(1,3)}$ and $(p-3)$-dimensional flat compact space (e.g. commutative torus $T^{(p-3)}$). The curvature term $R_{\it{E}}$ in (\ref{2}) being compactified from $D$ to 4 dimensions will give the conventional Einstein term of the action:

\begin{eqnarray}
\label{28}
&&M_{\rm{Pl}}^{2}\int{R^{(4)}\sqrt{-g^{(4)}}}\,d^{4}x= \nonumber \\
&&=M^{D-2}V_{(p-3)}\Omega_{n}\int_{0}^{z_{\it{vis}}}b^{p+1}Na^{n}\left(\frac{1}{b^{2}}\int{R^{(4)}\sqrt{-g^{(4)}}}\,d^{4}x\right) \,dz,
\end{eqnarray}
where $V_{(p-3)}$ - volume of commutative torus $T^{(p-3)}$ at $z=0$; $\Omega_{n}$ - volume of unit $n$-sphere. 

The substitution of $a$, $b$, $N$ from (\ref{18}) with account of (\ref{19}), (\ref{21}) and integration over $z$ in (\ref{28}) gives:

\begin{equation}
\label{29}
\frac{M_{\rm{Pl}}^{2}}{M^{2}}\cong(M^{p-3}V_{(p-3)})\left(\frac{M}{\sigma}\right)^{n+1}\Omega_{n}K_{1}^{n+1}\frac{2(p+1)(n-1)+(p+n)\alpha^{2}}{2(p-1)(n-1)^{2}+(n+1)(p+n)\alpha^{2}},
\end{equation}
we put down here only main term received by replacing $z_{\it{vis}}$ by $L$ in the upper limit of integration over $z$ in (\ref{28}); $K_{1}$ is defined in (\ref{21}).

Following general Randall and Sundrum approach~\cite{Randall} it is supposed that fundamental string scale $M$ is also the fundamental scale of action of matter trapped at the "visible" brane. Hence the observed mass scale $m$ of the visible matter is given by the value of warp factor $b$ in metric (\ref{4}) at the position of the "visible" brane:

\begin{equation}
\label{30}
\frac{m}{M}=\frac{b(z_{\it{vis}})}{b(0)}=\left(1-\frac{z_{\it{vis}}}{L}\right)^{\xi},
\end{equation}
see (\ref{18}), (\ref{19}) for $b$, $\xi$ and (\ref{26}), (\ref{27}) for two considered options of values of $z_{\it{vis}}$.

Mass hierarchy $m/M_{\rm{Pl}}$ which we are looking for is a product of two small numbers $m/M$ and $M/M_{\rm{Pl}}$ calculated from (\ref{30}) and (\ref{29}) correspondingly:

\begin{equation}
\label{31}
\frac{m}{M_{\rm{Pl}}}=\frac{m}{M}\,\frac{M}{M_{\rm{Pl}}}.
\end{equation}

To estimate the result quantitatively we shall suppose that "input" dimensional parameters in the RHS of (\ref{29}) and (\ref{30}) (with account of (\ref{26}) or (\ref{27}) for $z_{\it{vis}}$) are of one and the same order, i.e.:

\begin{equation}
\label{32}
\frac{\sigma}{M}=1, \qquad  M^{p-3}V_{(p-3)}=1. 
\end{equation}

With account of (\ref{32}) results of calculations of mass hierarchy (\ref{31}) and of ratio of string scale to Planck scale are presented for four particular cases (values of $\alpha$ are taken from (\ref{3}), $m^{(\it{st})}$ is received when condition (\ref{11}) for "string" curvature is used, $m^{(\it{E})}$ follows from condition (\ref{12}) for "Einstein" curvature):

(I) $D$=10, $p$=4, $n$=4, $\alpha=\frac{1}{2}$:

\begin{equation}
\label{33}
\frac{M}{M_{\rm{Pl}}}=4 \cdot 10^{-4}, \qquad \frac{m^{(\it{st})}}{M_{\rm{Pl}}}=1 \cdot 10^{-6}, \qquad \frac{m^{(\it{E})}}{M_{\rm{Pl}}}=9 \cdot 10^{-14};
\end{equation}

(II) $D$=10, $p$=5, $n$=3, $\alpha=1$:

\begin{equation}
\label{34}
\frac{M}{M_{\rm{Pl}}}=5 \cdot 10^{-3}, \qquad \frac{m^{(\it{st})}}{M_{\rm{Pl}}}=2 \cdot 10^{-3}, \qquad \frac{m^{(\it{E})}}{M_{\rm{Pl}}}=8 \cdot 10^{-4};
\end{equation}

(III) $D$=11, $p$=4, $n$=5, $\alpha=\frac{1}{3\sqrt{2}}$:

\begin{equation}
\label{35}
\frac{M}{M_{\rm{Pl}}}=3 \cdot 10^{-5}, \qquad \frac{m^{(\it{st})}}{M_{\rm{Pl}}}=8 \cdot 10^{-15}, \qquad \frac{m^{(\it{E})}}{M_{\rm{Pl}}}=2 \cdot 10^{-91};
\end{equation}

(IV) $D$=9, $p$=3, $n$=4, $\alpha=\frac{1}{\sqrt{14}}$:

\begin{equation}
\label{36}
\frac{M}{M_{\rm{Pl}}}=3 \cdot 10^{-4}, \qquad \frac{m^{(\it{st})}}{M_{\rm{Pl}}}=2 \cdot 10^{-10}, \qquad \frac{m^{(\it{E})}}{M_{\rm{Pl}}}=1 \cdot 10^{-47}.
\end{equation}

\section{Discussion}

\qquad Strong, non-analytical dependence of ratio $m/M$ (\ref{30}) on dimensionalities and dilaton coupling constant $\alpha$ (exponent $\xi \sim \alpha^{-2}$ etc.) is perhaps the main feature of the proposed approach. Because of it the appearance of ridiculously small numbers (like $10^{-91}$ in (\ref{35})) is not a surprise. It must be noted however that when physically more grounded restriction (\ref{11}) for curvature in "string" metric is used the values of mass hierarchy $m^{(\it{st})}/M$ look more sensible, in any case in the examples considered in previous Section.

Because of high sensibility of the result to the values of "input" parameters the main problem of the approach is the arbitrary conditions (\ref{32}) for brane's tension $\sigma$ and for volume of commutative torus (in case $p>3$). The small change of ratio $\sigma / M$ will drastically influence the "predicted" value of mass hierarchy. Although the hypotheses that value of brane's tension is of order of fundamental scale $M$ seems quite natural the proposed approach can claim to be predictive only in case the brane action (\ref{7}) will be deduced from first principles.

It is interesting to note that for special choice of dilaton coupling constant 

\begin{equation}
\label{37}
\alpha^{2}=2(n-1)^{2}(p+n)^{-1}
\end{equation}
fine-tuning condition (\ref{22}) gives $\kappa = 0$. For this choice of $\alpha$ there is no need in the induced curvature term in brane action (\ref{7}) since in this case dependence of metric scales $a$ and $b$ on $z$ is one and the same ($\xi=1$ in (\ref{18})) and corresponding Israel conditions coincide. For the "string" values of $\alpha$ given in (\ref{3}) the choice (\ref{37}) is possible if dimensionalities of subspaces $p$, $n$ satisfy the relation:

\begin{equation}
\label{38}
p=3n-4.
\end{equation}
This is fulfilled in particular in most familiar version of superstring action ($D=10$, $n=3$, $p=5$). This case however is not too interesting from the point of view of calculation of large mass hierarchy (see (\ref{34})). Of course it is possible to get large numbers by increasing the number of dimensions of space-time. E.g. (\ref{37}), (\ref{38}) are satisfied by the choice $D=26$, $n=7$, $p=17$, in this case large mass hierarchy results from big value of $M_{\rm{Pl}}/M$ given by (\ref{29}); the result however crucially depends on the choice of the volume $V_{(14)}$ of 14-dimensional commutative torus. To fix the moduli of the target space, to find physical grounds for this fixing, is a longstanding problem of string theory.

It would also be interesting if exact solution in the theory given by the action (\ref{2}) was found for metric of type (\ref{4}) with non-flat space-time $M^{(p+1)}$ in general case $\alpha \ne 0$. It may provide a tool for calculation of observed small cosmological constant (cf.~\cite{Alt}).

\section*{Acknowledgements} Author is grateful for comments and criticism  to participants of Quantum Field Theory Seminar of the Theoretical Physics Department, Lebedev Physical Institute. This work was partially supported by the grant LSS-1578.2003.2

\end{document}